\documentclass[aps,prd,11pt,superscriptaddress,preprintnumbers,nofootinbib]{revtex4-1}
\usepackage{amsmath,bm}
\usepackage{physics}
\usepackage{graphicx}
\usepackage{amssymb}
\usepackage[colorlinks, citecolor=magenta, urlcolor=blue]{hyperref}
\usepackage[normalem]{ulem}

\newcommand{\beq}[1][]{\begin{equation}\label{#1}}
\newcommand{\eeq}{\end{equation}}
\newcommand{\bse}{\begin{subequations}}
	\newcommand{\ese}{\end{subequations}}

\usepackage{color}
\definecolor{darkmagenta}{rgb}{0.4,0,0.4}
\definecolor{darkgreen}{rgb}{0.0, 0.5, 0.0}

\begin{document}

\preprint{{\vbox{\hbox{\bf MSUHEP-23-020}}}}
\vspace*{0.2cm}
	
\title{Probing Parton distribution functions at large $x$ via Drell-Yan Forward-Backward Asymmetry}
	
\author{Yao Fu}
\email{yfu@mail.ustc.edu.cn}
\affiliation{Department of Modern Physics, University of Science and Technology of China, Jinzhai Road 96, Hefei, Anhui, 230026, China}
	
	\author{Raymond Brock}
	\email{brockr@msu.edu}
	\affiliation{Department of Physics and Astronomy,
		Michigan State University, East Lansing, Michigan 48824, USA}

	\author{Daniel Hayden}
\email{haydend5@msu.edu}
\affiliation{Department of Physics and Astronomy,
	Michigan State University, East Lansing, Michigan 48824, USA}
	
	\author{C.-P. Yuan}
	\email{yuanch@msu.edu}
	\affiliation{Department of Physics and Astronomy,
		Michigan State University, East Lansing, Michigan 48824, USA}
	
	\date{\today}
	
	\begin{abstract}

The forward-backward asymmetry of the Drell-Yan process in dilepton decays at high invariant masses can be used to probe the parton distribution functions at large $x$. The behavior of three modern PDF sets (CT18NNLO, MSHT20, and NNPDF4.0) are compared, and updated under various scenarios via {\tt ePump} using proton-proton collision pseudo-data generated at $\sqrt{s}$ = 13~TeV with 3000~${\rm fb}^{-1}$ of integrated luminosity.

	\end{abstract}

\maketitle

\section{Introduction}

\noindent Many of the future, high luminosity, precision measurements at the Large Hadron Collider (LHC) will be limited by modeling and theoretical uncertainties. Prominent among these are uncertainties related to knowledge of parton distribution functions (PDFs). Kinematics of the wide range of measurements and searches at the LHC span parton $x$ ranges from very low where gluon and antiquark densities are relevant to high $x$ regions dominated by valance quark densities. An intriguing approach to further limiting PDF uncertainties might be to, rather than rely on global, all-purpose PDF sets, create targeted "boutique" PDF sets designed for specific purposes. Judicious kinematic selection of LHC data as strategic, supplementary inputs to PDF fitting might lead to legitimate reductions from such special-purpose PDF sets.  Neutral-current  Drell-Yan (NCDY) backgrounds constitute a trial of this idea as high mass resonant and non-resonant searches are standard targets for new physics searches but they are now compromised by PDF uncertainties, and will be more-so in the future.  Standard model (SM) backgrounds that must be modeled very well are experimentally uncomplicated and amenable to high experimental precision. So NCDY studies seem a good way to  explore this idea of specialized-PDFs. 

We tested that approach in a previous study~\cite{Willis:2018yln} and demonstrated  how measurements of  NCDY below $m_{\ell\ell}$ = 1~TeV, bring significant improvement to high-mass PDF uncertainties. That paper studied the triple differential cross-section in the variables $m_{\ell\ell}$, $\cos(\theta^*)$, and $\eta_{\ell\ell}$ and simulated the sensitivity of modified PDFs in high mass searches for new physics. In a similar vein, this paper specifically looks at the behavior of various PDF sets in the description of the forward-backward asymmetry ($A_{FB}$) at high invariant masses. The {\tt ePump} Package~\cite{Schmidt:2018hvu} is then used to understand the effect of using these updated PDF nominal values and uncertainties if nature behaved according to the other PDFs considered. We note that others have begun exploring ways to improve NCDY PDF uncertainties at high invariant mass as well \cite{Fiaschi:2022wgl, Fiaschi:2021sin, Fiaschi:2021okg, Amoroso:2020fjw, Accomando:2019vqt, Accomando:2018nig, Accomando:2017scx}.

The following equation describes the relationship between initial state parton momentum fractions and kinematic variables in hadron collisions:

\begin{eqnarray}
	x_1, x_2 = \frac{Q}{\sqrt{s}}e^{\pm y},
 \label{partons}
\end{eqnarray}

\noindent where $x_1$, $x_2$ are the momentum fractions of partons in the incoming beam, $Q$ is the energy scale, $\sqrt{s}$ is the collision energy, and $y$ is the rapidity of the system. According to Equation~\eqref{partons}, high mass events directly correspond to the high $x$ region in PDFs. In modern PDFs, the valence quarks are well-known for $x \sim 0.3$ while the sea quark PDFs, especially the strangeness PDFs, are hardly known for $x > 0.3$. However, for $x > 0.7$ even the valence PDFs are not determined well. Kinematic distributions for the high mass Drell-Yan process, especially the forward-backward asymmetry ($A_{FB}$) directly provide valence parton information in that large $x$ region. Therefore, $A_{FB}$ measurements in the high mass region at the LHC can provide complementary information on large $x$ PDFs to the science program at the Electron-Ion Collider (EIC)~\cite{AbdulKhalek:2022hcn}.

In this study, three modern PDF sets are compared: CT18NNLO~\cite{Hou:2019efy}, MSHT20~\cite{Bailey:2020ooq}, and NNPDF4.0~\cite{NNPDF:2021njg} by generating proton-proton ($pp$) collision pseudo-data corresponding to $\sqrt{s}$ = 13~TeV and an integrated luminosity of 3000~${\rm fb}^{-1}$. These pseudo-data are used as input to the {\tt ePump} Package to update either CT18NNLO or MSHT20 as the underlying theory input. By updating the PDFs, both the central value of the PDF, and the 68\% confidence level (C.L.) variation uncertainty (i.e. for CT18NNLO this uncertainty describes the variation of the underlying eigen-vectors) will change. This study also carefully considers the role of the tolerance used in {\tt ePump} when updating PDF sets with new data, and discusses how to appropriately set these parameters to obtain a realistic result.

\section{Forward-backward asymmetry($A_{FB}$) and Dilution factor}

\begin{figure}[htbp]
	\centering
	\includegraphics[width=0.47\textwidth]{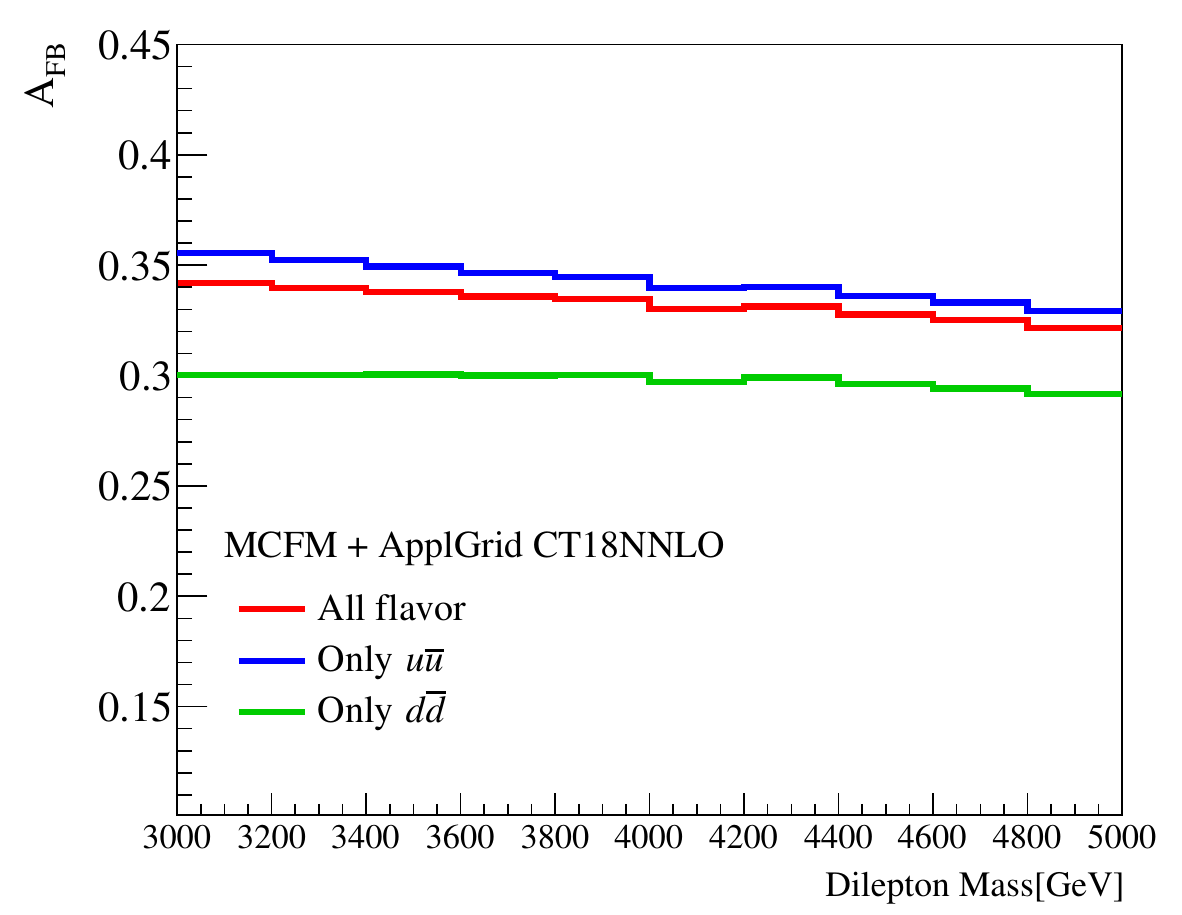}
	\includegraphics[width=0.47\textwidth]{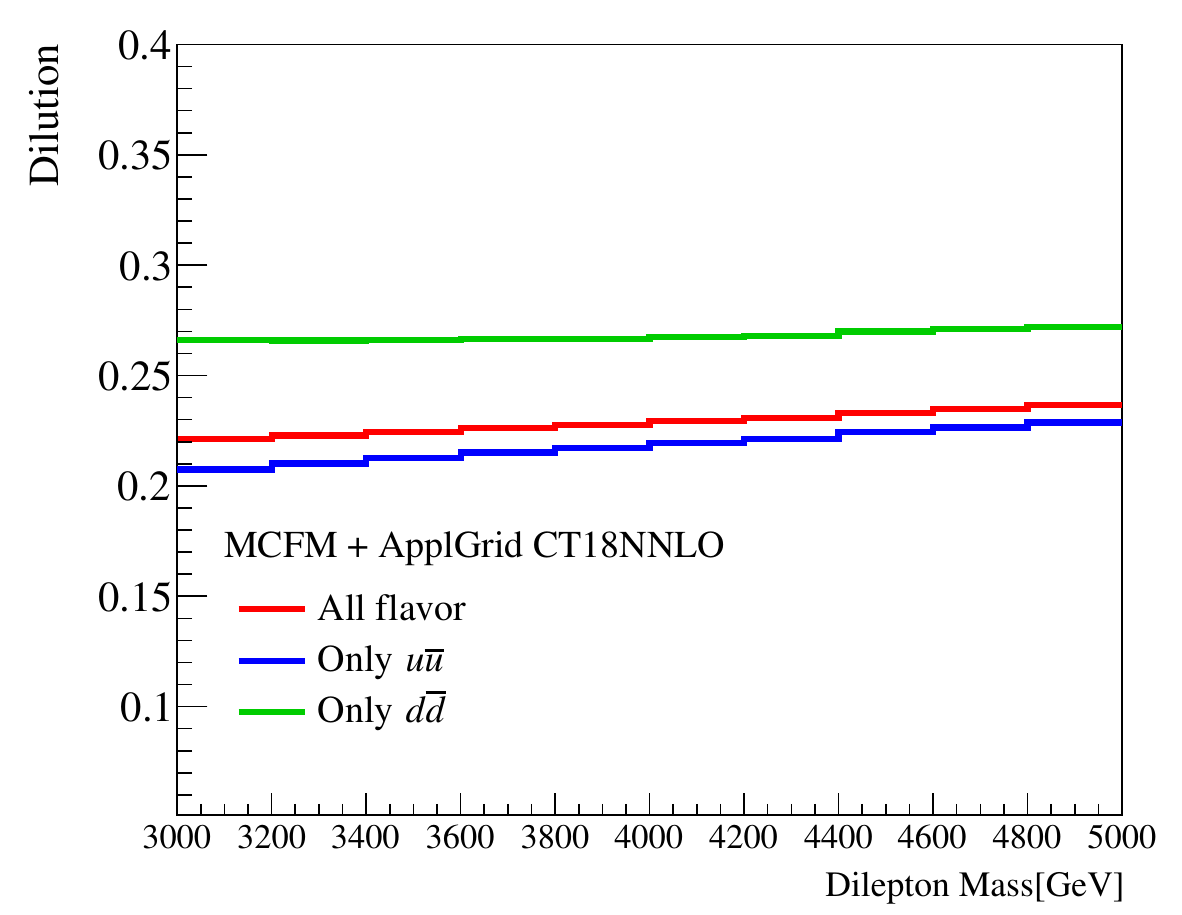}
	\caption{The $A_{FB}$ (left) and Dilution (right) spectrum as a function of dilepton mass for MCFM at NLO accuracy in the high mass region ($M_{ll} > 3000$~GeV), using the CT18NNLO PDF set. The red, blue, and green curves represents the all flavor combination, only $u\bar{u}$ contribution, and only $d\bar{d}$ contribution, respectively.}
	\label{fig:HighMassAFB_flavor}
\end{figure}

$A_{FB}$ is defined as,
\begin{eqnarray}
	A_{FB} = \frac{N_F - N_B}{N_F + N_B}\,,
 \label{afb}
\end{eqnarray}
where $N_F$ and $N_B$ are the number of forward and backward events, respectively. 
In the NCDY process, the scattering angle $\theta$ is defined by the momentum direction of the outgoing fermion $f_j$ relative to the momentum direction of the incoming fermion $f_i$. The sign of $\cos\theta$ is used to define forward ($\cos\theta>0$) and backward ($\cos\theta<0$) events.
At hadron colliders, due to the unknown momentum direction of the incoming quarks, forward and backward events are defined in the Collins-Soper (CS) rest frame~\cite{PhysRevD.16.2219}, where the polar and azimuthal angles are defined relative to the two hadron beam directions. The $z$ axis is defined in the rest frame of the DY pair, bisecting the angle between the incoming hadron momentum and the negative of the other hadron momentum.
The cosine of the polar angle $\theta^{*}$ between the momentum direction of the outgoing lepton $l^{-}$ and the $\hat{z}$ axis in the CS frame is defined as the scattering angle of the DY pair at hadron colliders, which can be calculated directly from the laboratory frame lepton quantities by

\begin{eqnarray}
	\cos\theta_{CS}^{*} = c \, \frac{2(p^{+}_{1}p^{-}_{2} -
		p^{-}_{1}p^{+}_{2})}
	{m_{ll}\sqrt{m^2_{ll} + p^{2}_{{\rm T},ll}}} \,,
\end{eqnarray}
where the scalar factor $c$ (either 1 or -1) is defined for the Tevatron and the LHC, respectively, as 
\begin{eqnarray}
	c =\Bigl\{
	\begin{array}{ll}
		1, &   \mbox{for the Tevatron (a proton-antiproton collider)} \\
		\vec{p}_{Z,ll}/|\vec{p}_{Z,ll}|, & \mbox{for the LHC (a proton-proton collider)}\,. \\
	\end{array}
	\label{eq:zdirection}
\end{eqnarray}
And thus, the sign of the $z$ axis is defined as the proton beam direction for the Tevatron, and as the sign of the boost direction of the lepton pair with respect to the $z$ axis in the laboratory frame on an event-by-event basis for the LHC.
The variables $p_{Z,ll}$, $m_{ll}$, and $p_{{\rm T},ll}$ denote the longitudinal momentum, invariant mass and transverse momentum of the dilepton system, 
respectively, and,
\begin{eqnarray}
	p^{\pm}_i = \frac{1}{\sqrt{2}}(E_i \pm p_{Z,i}) \,,
\end{eqnarray}
where the lepton (anti-lepton) energy and longitudinal momentum are $E_{1}$ and $p_{Z,1}$ ($E_{2}$ and $p_{Z,2}$), respectively. 
DY events are therefore defined as forward ($\cos\theta_{CS}^{*}>0$) or backward ($\cos\theta_{CS}^{*}<0$) according to the direction of 
the outgoing lepton in this frame of reference.
In the case of the LHC, one can define another frame such that the $z$ axis is oriented to the quark direction of motion. We use $\cos\theta_{q}^{*}$ to denote this case, and define a coefficient $c$ as,
\begin{eqnarray}
	c = \vec{p_{q}}/|\vec{p_{q}}|,
\end{eqnarray}
We then use $\cos\theta_{h}^{*}$ to denote the case using the lepton pair momentum to define the $z$ axis.
One can easily obtain the relationship between $\cos\theta_{h}^{*}$ and $\cos\theta_{q}^{*}$,
\begin{eqnarray}
	\begin{array}{ll}
		\cos\theta_{h}^{*} = \cos\theta_{q}^{*}, & {\rm for} \, \mbox{$E(q)>E(\bar{q})$} \\
		\cos\theta_{h}^{*} = -\cos\theta_{q}^{*}, & {\rm for} \, \mbox{$E(q)<E(\bar{q})$}. \\
	\end{array}
\end{eqnarray}
As introduced in previous studies~\cite{Fu:2020mxl,Yang:2021cpd,PhysRevD.106.033001}, the dilution factor ($D$) quantifies the probability that the energy of the anti-quark is larger than the energy of quark. 
When the quark carries higher energy, the number of forward and backward events in the two different frames will be the same, which has a probability of $(1-D)$. When the antiquark carries higher energy, the number of forward and backward events in the two different frames will have a different sign, which has a probability of $D$. Finally, the number of forward and backward events $N_F^h$ and $N_B^h$ defined by the $\cos\theta_{h}^{*}$ can be written as
\begin{eqnarray}
	\begin{array}{ll}
            N_F^h=(1-D)N_F^q+D N_B^q, \\
            N_B^h=(1-D)N_B^q+D N_F^q,
        \end{array}
\end{eqnarray}
\noindent where the $N_F^q$ and $N_B^q$ represent the number of forward and backward events defined by the $\cos\theta_{q}^{*}$.
As a result, the relationship between the $A_{FB}$ defined by the $\cos\theta_{h}^{*}$, and the $A_{FB}$ defined by the $\cos\theta_{q}^{*}$, can be roughly written as,

\begin{eqnarray}\label{eq:dilution}
	A_{FB}^{h} \approx (1-2D)A_{FB}^{q},
\end{eqnarray}

\noindent where $A_{FB}^h$ and $A_{FB}^{q}$ are the asymmetries defined by $\cos\theta_{h}^{*}$ and $\cos\theta_{q}^{*}$, respectively.
Writing this for other flavors, Equation~\eqref{eq:dilution} can be written precisely as

\begin{eqnarray}
    A_{FB}^h = \sum_{f} \frac{N_f}{N} (1-2D_f) A_{FB}^f,
\end{eqnarray}

where $f$ represents the index of flavors which is directly coupled with $Z$ boson. Here the $f$ contains five flavors: $u\bar{u}$, $d\bar{d}$, $s\bar{s}$, $c\bar{c}$, and $b\bar{b}$. The $N$ and $N_f$ represent the total number of events, and the number of events with a specific flavor, respectively. $D_f$ represents the dilution factor in the process with a certain flavor, and $A_{FB}^f$ is defined by the $\cos\theta_{q}^{*}$ with flavor dependence. The $up$-type and $down$-type flavor coupled with $Z$ boson have different $A_{FB}$ values, where a more detailed discussion can be found in the $A_{FB}$ factorization study~\cite{PhysRevD.106.033001}. The dilution factor $D$ defined in Equation~\eqref{eq:dilution} can be roughly treated as an average of all the flavor combinations.  
Fig.~\ref{fig:HighMassAFB_flavor} shows the $A_{FB}$ and Dilution factor as a function of dilepton mass for all flavor combinations, $D$, only $u\bar{u}$ contributions, $D_u$, and only $d\bar{d}$ contributions, $D_d$, for the CT18 PDF. The calculation is done by using the MCFM program~\cite{Campbell:2019dru} at NLO, interfaced to APPLgrid~\cite{Carli:2010rw}. The $s$, $c$, and $b$ quark contribution are about the same as their antiquarks in CT18 PDFs, so the dilution factor $D_s$, $D_c$, and $D_b$ will be $0.5$, and the $A_{FB}$ defined by $\cos\theta_{h}^{*}$ for those processes will be zero. 
For CT18NNLO, ${\bar s}(x,Q_0)$ is assumed to be identical to $s(x,Q_0)$ at $Q_0=1.3$ GeV. When the PDFs are evolved to higher energy scale $Q$, ${\bar s}(x,Q_0)$ and $s(x,Q_0)$ can differ slightly at NNLO and beyond. 
However, even in the case of allowing ${\bar s}(x,Q_0)$ not equal to $s(x,Q_0)$, such as in CT18As~\cite{Hou:2022onq}, the current constraint on the strangeness asymmetry is such that 
the dilution factor originating from the $s\bar{s}$ production channel will also be small. As shown in Fig.~\ref{fig:HighMassAFB_flavor} (right), the Dilution factor of the $u\bar{u}$ process is smaller than that of the $d\bar{d}$ process, while the $A_{FB}^h$ of the $u\bar{u}$ process is larger than the $d\bar{d}$ process.

\begin{figure}
    \centering
    \includegraphics[width=0.47\textwidth]{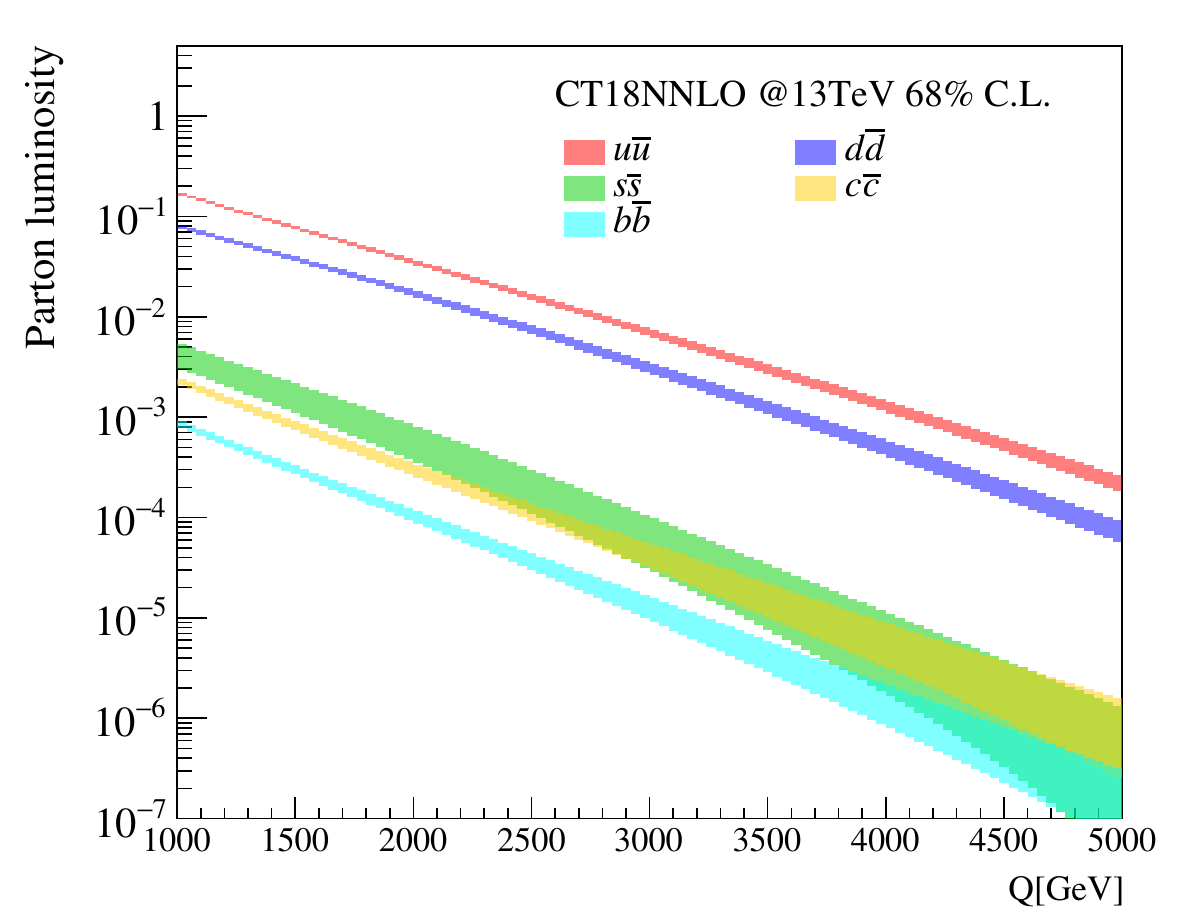}
    \caption{Parton luminosities of CT18NNLO at high invariant mass region for the initial state of $u\bar{u}$, $d\bar{d}$, $s\bar{s}$, $c\bar{c}$, and $b\bar{b}$. The width of the curves represents the PDF uncertainty for $68$\% C.L.}
    \label{fig:PDFLumi}
\end{figure}

Fig.~\ref{fig:PDFLumi} shows the parton luminosities for $u\bar{u}$, $d\bar{d}$, $s\bar{s}$, $c\bar{c}$, and $b\bar{b}$ processes. This shows that the $u\bar{u}$ process is dominant in the high invariant mass region. As a result, the $A_{FB}^h$ value of the all flavor combination is closer to the $A_{FB}^h$ in the $u\bar{u}$ process.

To specify what kind of flavor and $x$ region can be constrained, we write the dilution factor in parton language. According to the definition of the dilution factor, the equivalent in parton language can be written as,

\begin{eqnarray}
	D_q=\frac{f_{\bar{q}}(x_{1})f_{q}(x_{2})}{f_{q}(x_{1})f_{\bar{q}}(x_{2})+f_{\bar{q}}(x_{1})f_{q}(x_{2})}, & \mbox{\rm with  } x_{1}>x_{2}.
	\label{eq:dilution_parton}
\end{eqnarray}

\begin{figure}
    \centering
    \includegraphics[width=0.47\textwidth]{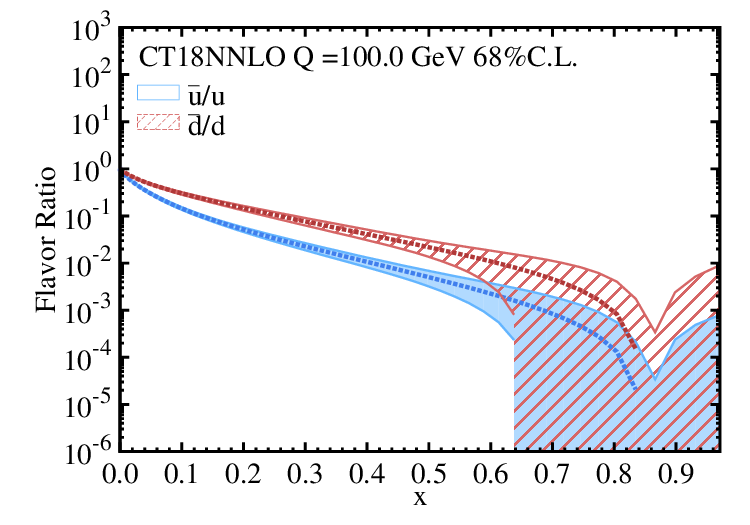}
    \caption{Central value and uncertainty of CT18 of $\bar{u}/u$ and $\bar{d}/d$ PDFs.}
    \label{fig:qboq}
\end{figure}

From this one can easily see that the dilution factor is sensitive to the relative difference between quarks and anti-quarks. 
Fig.~\ref{fig:qboq} shows the comparison of $\bar{u}/u$ and $\bar{d}/d$ quark PDFs for the CT18 PDF. Above $x>0.6$ the  relative difference between $\bar{u}$ and $u$ is larger than the difference between $\bar{d}$ and $d$, which 
explains the difference of $D_u$ and $D_d$ shown in Fig.~\ref{fig:HighMassAFB_flavor}.

This separation between  $\bar{u}$ and $u$ and  $\bar{d}$ and $d$  suggest that $A_{FB}$ in the high invariant mass region can provide sensitive information about the relative difference between quarks and antiquarks, namely, the valence information in the large $x$ region, which is not currently constrained by existing measurements at the LHC.

\section{PDF-updating using $A_{FB}$ in the high mass region}

In this section, we quantitatively show the impact of using $A_{FB}$ in the high invariant mass region, to update the PDFs under study and their associated uncertainties, using the \texttt{{\tt ePump}} package. \texttt{{\tt ePump}} is meant to  be used as a tool to approximate an update to a given PDF set and its respective PDF uncertainties in response to new kinds of input data, in this study, $A_{FB}$. Should an \texttt{{\tt ePump}} exercise suggest that these new kinds of data might inform the central PDF and/or reduce PDF uncertainties, then these new data types should be considered as additions to complete, global fitting.
In this study pseudo-data were generated for $A_{FB}$ in the high invariant mass region using ResBos~\cite{Balazs:1997xd,Isaacson:2017hgb} at N$^{3}$LL+NNLO in QCD at $\sqrt{s}$ = 13 TeV. Each PDF under study was used to generate the theory template,  required by {\tt ePump} to perform the update.  Electroweak parameters can affect  $A_{FB}$ distributions, but a previous study~\cite{Fu:2020mxl} shows that this mainly affects the $Z$ peak region. So we leave the electroweak parameters  set to be the same between pseudo-data and theory templates. 
In order to study the impact of various possible datasets, CT18, MSHT20, and NNPDF4.0 were each used to generate pseudo-data, which are then subsequently used to update the nominal PDF under study.

For the pseudo-data, samples were generated corresponding to an integrated luminosity of $3000$ fb$^{-1}$ in order to study the impact of data in future at the high luminosity LHC. No kinematic cuts are imposed in this study. $A_{FB}$ in the invariant mass region from 500-5000 GeV were used to perform the PDF updating (using 25 mass-bins of varying size to preserve good statistical precision).

\begin{figure}[htbp]
	\centering
	\includegraphics[width=0.47\textwidth]{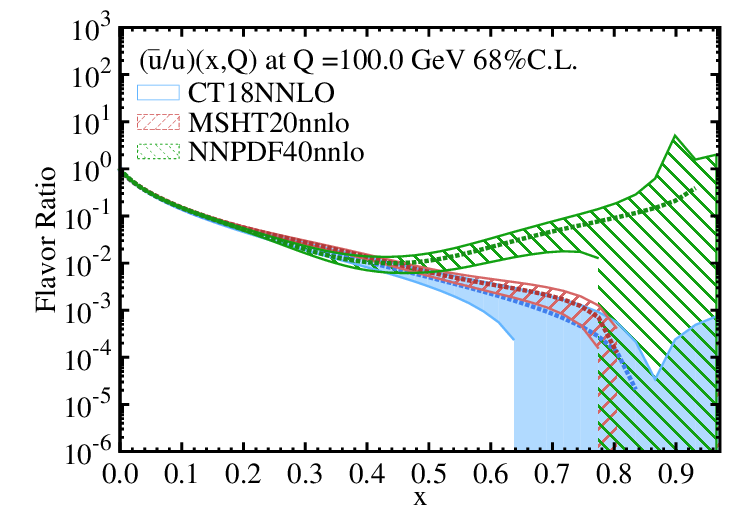}
	\includegraphics[width=0.47\textwidth]{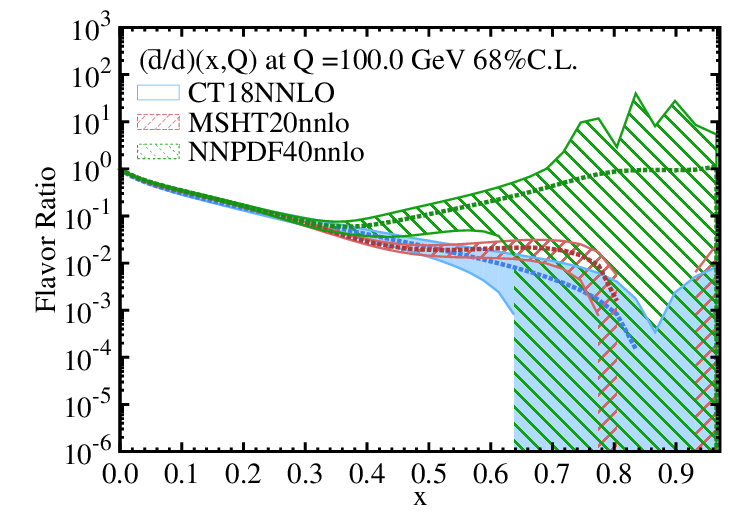}
	\includegraphics[width=0.47\textwidth]{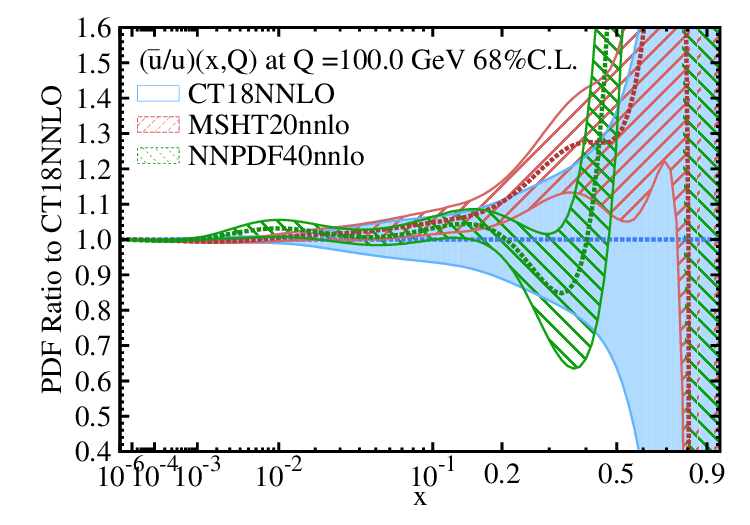}
	\includegraphics[width=0.47\textwidth]{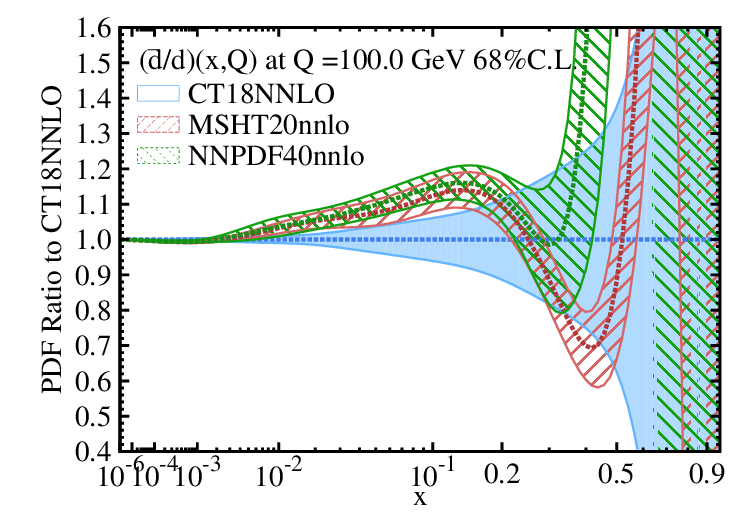}
	\caption{Central value and uncertainty of CT18, MSHT20, and NNPDF40 (top), and ratio (bottom) of the central value and uncertainty to the CT18NNLO central value of the $\bar{u}/u$ (left), and $\bar{d}/d$ (right) PDFs.}
	\label{fig:PDFs}
\end{figure}

As a starting point, Fig.~\ref{fig:PDFs} shows the central value and uncertainty of CT18, MSHT20, and NNPDF40, as well as their ratio compared to CT18, for $\bar{u}/u$, and $\bar{d}/d$. 

\subsection{PDF-updating for CT18}

\begin{figure}[htbp]
	\centering
	\includegraphics[width=0.47\textwidth]{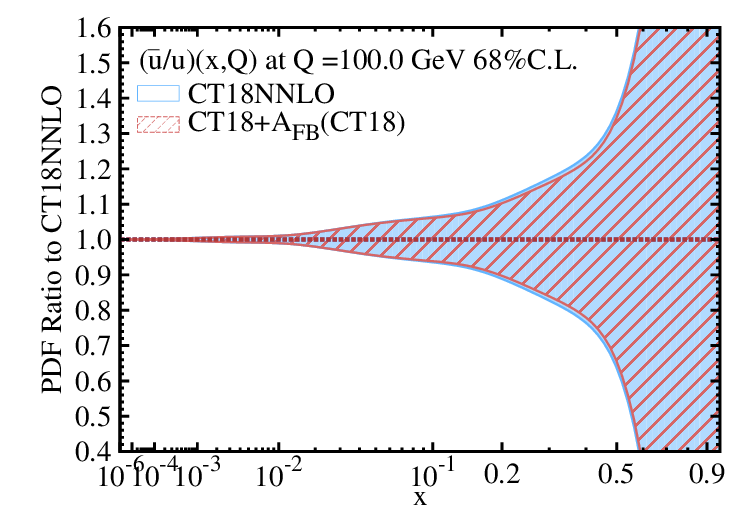}
	\includegraphics[width=0.47\textwidth]{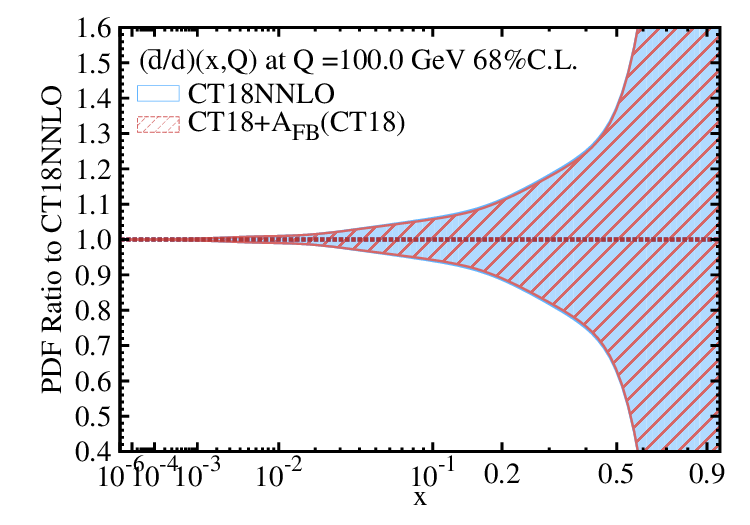}
	\caption{PDF update of CT18 for $\bar{u}/u$ (left), and $\bar{d}/d$ (right) using $A_{FB}$ pseudo-data generated using CT18. The central value and uncertainty are compared to the CT18 central value.}
	\label{fig:PDFupdate_CT18}
\end{figure}

Firstly, the CT18 PDF was updated using psuedo-data generated by CT18 itself, based on the flavor combination of $\bar{u}/u$, and $\bar{d}/d$. This is because according to the definition of the dilution factor in Equation~\eqref{eq:dilution_parton}, the most sensitive flavor combination of $A_{FB}$ data is $\bar{u}/u$, and $\bar{d}/d$. Fig.~\ref{fig:PDFupdate_CT18} shows the ratio of CT18 and its uncertainties before and after updating. Since CT18 was used for both the pseudo-data and the theory templates, the central values of the PDF set does not change, as expected. Similarly, the error band after the update shows only a slight reduction of a few percent compared to the original. This is due to  the relatively small number of events in the high invariant mass region. A similar study~\cite{Fiaschi:2021okg} which used \texttt{xFitter} instead of \texttt{{\tt ePump}} to update CT18 using $A_{FB}$ from the $Z$ peak region, found a relative improvement in the PDF uncertainty at $x$ = 10$^{-4}$ of $\sim$63\%. We also calculated the result of limiting our inputs to the $Z$ peak region instead of the high mass region, and found an improvement in the PDF uncertainty of $\sim$53\% for the same $x$ range. However, \texttt{xFitter} and \texttt{{\tt ePump}} use different tolerances to weight the importance of new data in the PDF update, which is an important topic handled in more detail in the Appendix of this paper. Fig.~\ref{fig:PDFupdate_MSHT20} then show the results of updating the CT18 PDF using the pseudo-data generated by MSHT20 and NNPDF4.0,  respectively. Due to the different $A_{FB}$ distributions predicted by MSHT20 and NNPDF4.0, the effects of new $A_{FB}$ data in the high mass region are absorbed into the $\bar{u}/u$, and $\bar{d}/d$ PDFs in the large $x$ region. The updated $\bar{u}/u$, and $\bar{d}/d$ PDFs show obvious deviations compared to the CT18 PDF.

\begin{figure}[htbp]
	\centering
	\includegraphics[width=0.47\textwidth]{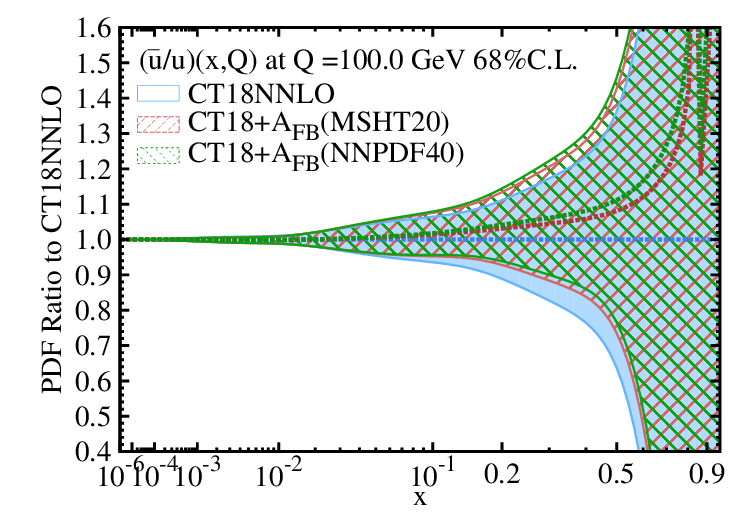}
	\includegraphics[width=0.47\textwidth]{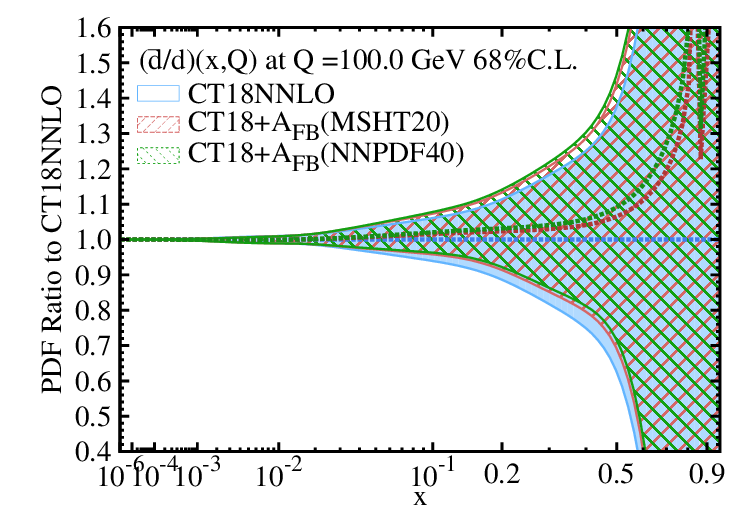}
	\caption{PDF update of CT18 for $\bar{u}/u$ (left), and $\bar{d}/d$ (right) using $A_{FB}$ pseudo-data generated using MSHT20 and NNPDF4.0. The central value and uncertainty are compared to the CT18 central value.}
	\label{fig:PDFupdate_MSHT20}
\end{figure}

\newpage

\subsection{PDF-updating for MSHT20}
\label{sec:update_msht}

In this section, another set of PDF updating results are presented based on using MSHT20 as the nominal PDF set. The pseudo-data are kept the same as the previous part of this study. The difference in this section is that the theory templates are generated using the MSHT20 PDF instead of CT18.
Firstly, in MSHT20 the nonperturbative shape parameters are different with respect to CT18, which leads to different behavior for $\bar{u}/u$ and $\bar{d}/d$, and means it is interesting to repeat the study using MSHT20 as the nominal PDF. Secondly, since the tolerance used in MSHT20 is about $10$, which is smaller than CT18 by a factor of $3$ (The average tolerance for CT18 is about $30$ for $68\%$ C.L.), the impact of new data on MSHT20 could be much stronger. A more in-depth discussion on the effect of using different tolerances is described in Appendix~\ref{app:tolerance}.
However, as shown in Fig.~\ref{fig:PDFUnc}, the PDF uncertainty of $A_{FB}$ for MSHT20 is much smaller than for CT18. The updating procedure also conveys such information, which when coupled with the smaller tolerance being used, finally ends up with a similar result in terms of relative change to the CT18 case.

\begin{figure}
	\centering
	\includegraphics[width=0.5\textwidth]{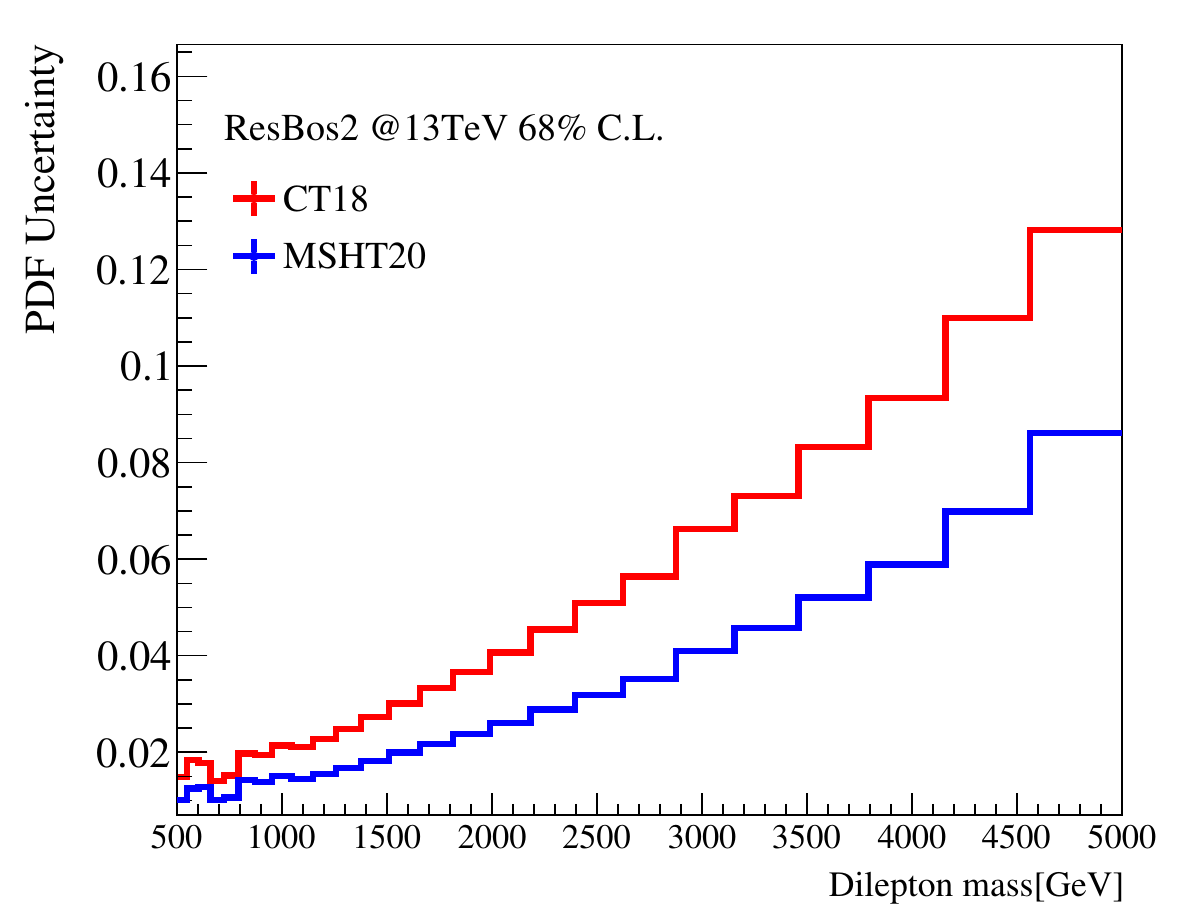}
	\caption{PDF uncertainty of $A_{FB}$ in high invariance mass region corresponding to $68\%$ C.L. for CT18 and MSHT20.}
	\label{fig:PDFUnc}
\end{figure}

\begin{figure}[htbp]
	\centering
	\includegraphics[width=0.47\textwidth]{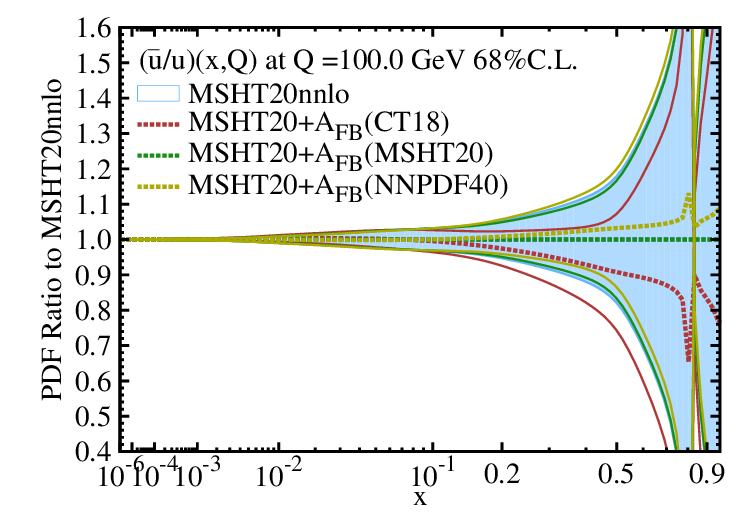}
	\includegraphics[width=0.47\textwidth]{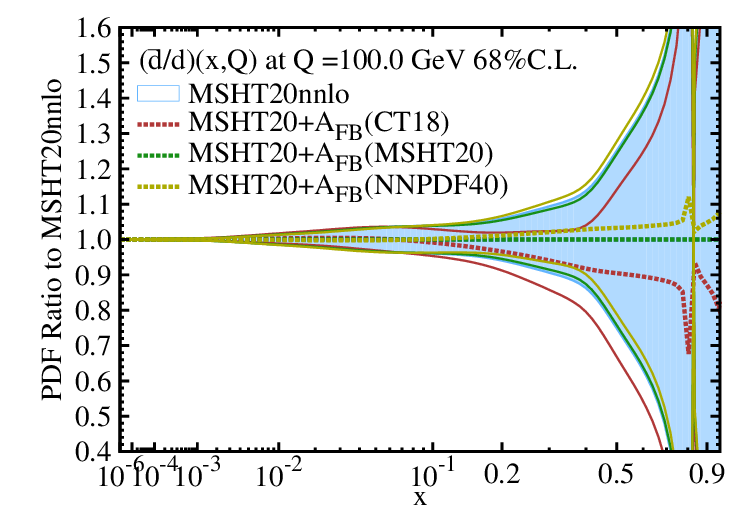}
	\caption{PDF update for $\bar{u}/u$ (left), and $\bar{d}/d$ (right) using $A_{FB}$ pseudo-data generated using CT18, MSHT20, and NNPDF40. The central value and uncertainty are compared to the MSHT20 central value.}
 	\label{fig:PDFupdate_CT18_ForMSHT}
\end{figure}

Fig.~\ref{fig:PDFupdate_CT18_ForMSHT} shows the PDF updating results when using the pseudo-data generated by CT18, MSHT20, and NNPDF40 to update MSHT20.

\clearpage

\section{PDF Comparisons}

\begin{figure}[htbp]
	\centering
	\includegraphics[width=0.47\textwidth]{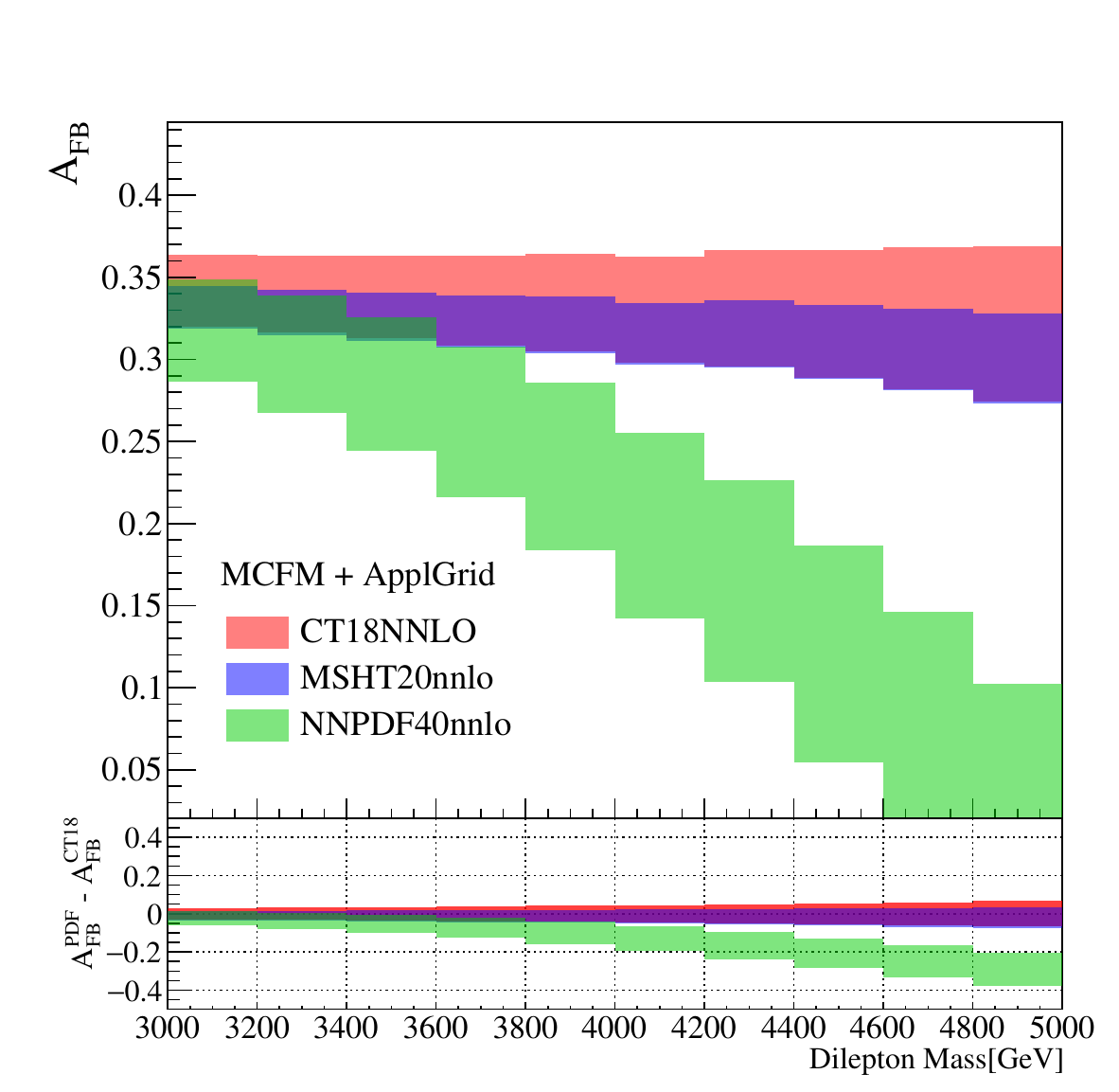}
	\includegraphics[width=0.47\textwidth]{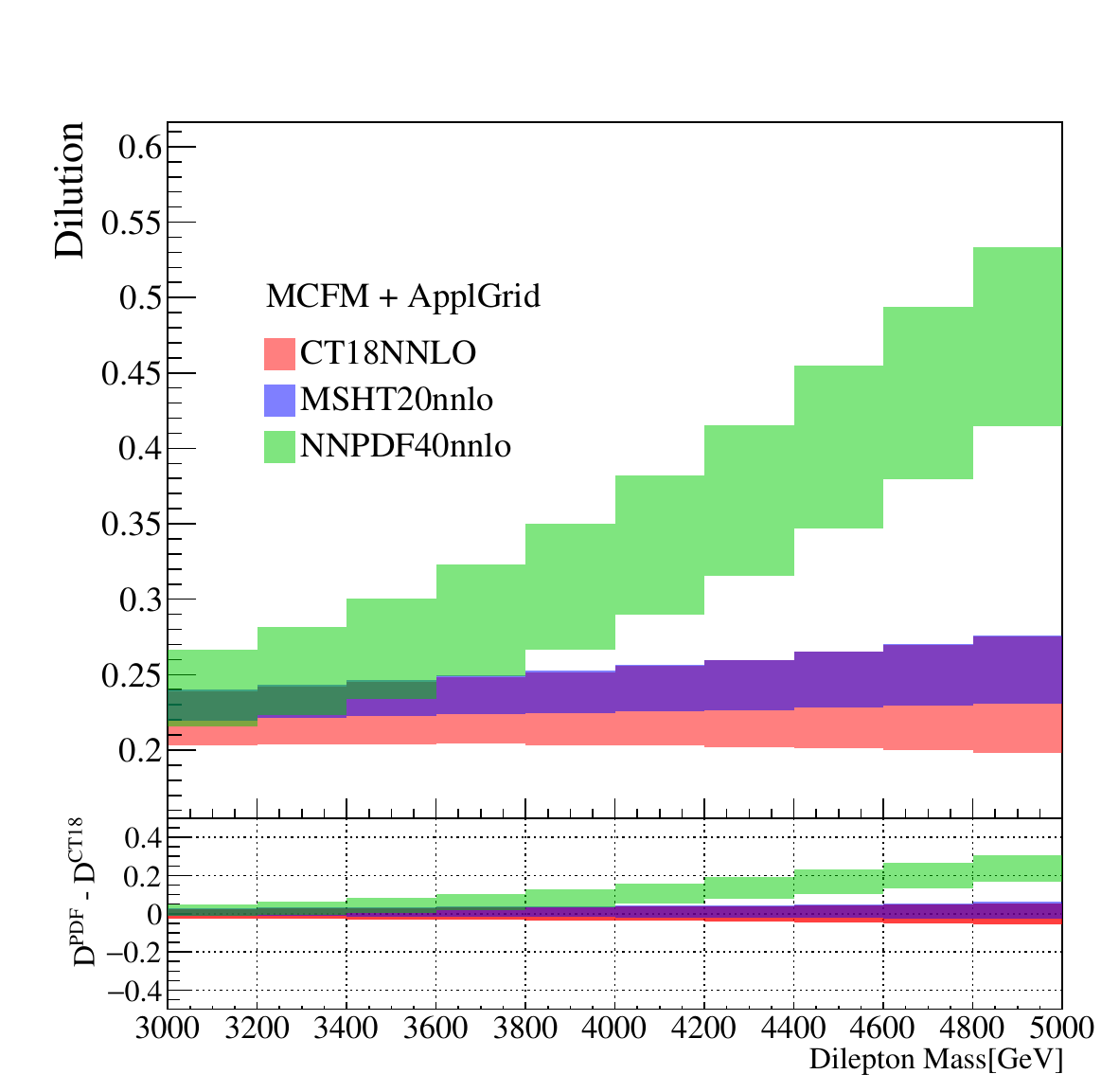}
	\caption{Comparison of the $A_{FB}$ (left) and Dilution (right) with different PDF input for MCFM at NLO accuracy in the high mass region ($M_{ll} > 3000$~GeV). The band represents the PDF uncertainty.}
	\label{fig:HighMassAFB}
\end{figure}

%It is clear from the results above, especially Figure~\ref{fig:PDFs}, that NNPDF4.0 behaves  differently than CT18 and MSHT20. 
Figure~\ref{fig:HighMassAFB} shows the MCFM~\cite{Campbell:2019dru} calculation at NLO for the $A_{FB}$ spectrum and the dilution factor as a function of dilepton invariant mass in the high invariant mass region with CT18, MSHT20, and NNPDF4.0 PDFs. Both Figures~\ref{fig:PDFs}  and \ref{fig:HighMassAFB} suggest  different behavior for NNPDF4.0 as compared with that of CT18 and MSHT20, and 
%In the high invariant mass region, $A_{FB}$ described by NNPDF4.0 drops off rapidly, and even becomes negative in the very high invariant mass region ($>$ 5000~GeV). In order to explain the negative $A_{FB}$ in very high invariant mass regions, a naive toy model has been discussed~\cite{Ball:2022qtp}. 
the dilution factor can explain the drop-off in $A_{FB}$ for NNPDF4.0 at high invariant masses compared to CT18 and MSHT20. Also, Figure~\ref{fig:HighMassAFB} (right) shows a growing feature for the dilution factor as a function of invariant mass for NNPDF4.0. According to Equation~\eqref{eq:dilution}, when the dilution factor $D$ is larger than 0.5, a negative $A_{FB}^{h}$ will be observed and that appears to be the case for NNPDF4.0 in the very high invariant mass region ($>$ 5000~GeV). In order to explain the negative $A_{FB}$ in very high invariant mass regions, a naive toy model has been discussed~\cite{Ball:2022qtp}.

Comparing $A_{FB}$ and the dilution factor described by CT18 and MSHT20 also reveals some differences in the methodology of the PDF global analysis, with the adoption of different datasets, higher-order theoretical calculations, and different choices of the non-perturbative parameterization forms of various parton PDFs.

\section{Conclusion}

This study presents the result of using the kinematic information from forward-backward asymmetry in the Drell-Yan process at high invariant mass to update various PDF sets (CT18NNLO, MSHT20, and NNPDF4.0). Pseudo-data were generated for $pp$ collisions at $\sqrt{s}$ = 13~TeV, for an integrated luminosity of 3000~${\rm fb}^{-1}$, using the separate PDFs, and input into {\tt ePump} to update the PDF central value and 68\% C.L. variation uncertainties for CT18NNLO and MSHT20. NNPDF4.0 was found to have a  different behavior in $A_{FB}$ versus dilepton invariant mass compared to the other PDFs considered. This is related to the dilution factor, and as such the most sensitive parts of the PDF to these effects are the ratio of $\bar{u}/u$ and $\bar{d}/d$. Using a dataset of 3000~${\rm fb}^{-1}$ appears to be sufficient to differentiate among the PDFs, such that if CT18NNLO or MSHT20 were assumed as the underlying theory, but nature behaved closer to NNPDF4.0, the former PDF sets would be updated by the new input data accordingly. While this new information from $A_{FB}$ at high invariant mass is effective at shifting the nominal prediction of PDFs, it was also shown that the effect on the PDF variation uncertainty is fairly modest. 

%into the kinematic regions that would aid most in reducing the PDF uncertainties for these relevant parts of the PDF could be informative. 

One could also make the argument against using high mass data in global PDF fits for the fear of somehow absorbing new physics into Standard Model (SM) PDFs. However, even low mass data could hide new physics in the tails of some distributions given enough data, and at some point SM-only PDFs would not be able to describe such a departure from the SM. Furthermore, given the different PDF 
behaviors observed in this study, the community should be aware that some PDFs (such as NNPDF4.0) might predict distributions of $A_{FB}$ versus dilepton invariant mass that are similar to many non-resonant new physics models (predicting high-mass drop off in $A_{FB}$). It should be noted that the choice of tolerance when using {\tt ePump} is important as it causes the new input data to have a stronger (for lower tolerances) or weaker (for higher tolerances) effect on the resulting PDF update. Choosing a tolerance that accurately reflects the new input data is vital to ensure a realistic resulting update.

Further studies are planned for specifically high-mass Drell-Yan searches to explore the notion that "boutique" PDFs created for specific uses might be developed using specified measurables to help constrain SM backgrounds in regions where, without more precise PDF modeling, new physics might lay hidden. Finally, we note that when real data become available, a full global fit has to be carried out to explore the need of new non-perturbative functional forms of the PDFs at the initial $Q_0$ scale to better describe the data, especially when the data provides new constraints on PDFs at very large or small $x$ regions. 

\section*{Acknowledgments}

We thank Liang Han and Siqi Yang for discussions. 
This work is partially supported by the U.S. National Science Foundation under Grant No. PHY-2013791, and PHY-2111226. C.P. Yuan is also grateful for the support from the Wu-Ki Tung endowed chair in particle physics.

\bibliography{reference}

\appendix
\section{Tolerance dependence in PDF updating}
\label{app:tolerance}

\begin{figure}[htbp]
	\centering
	\includegraphics[width=0.47\textwidth]{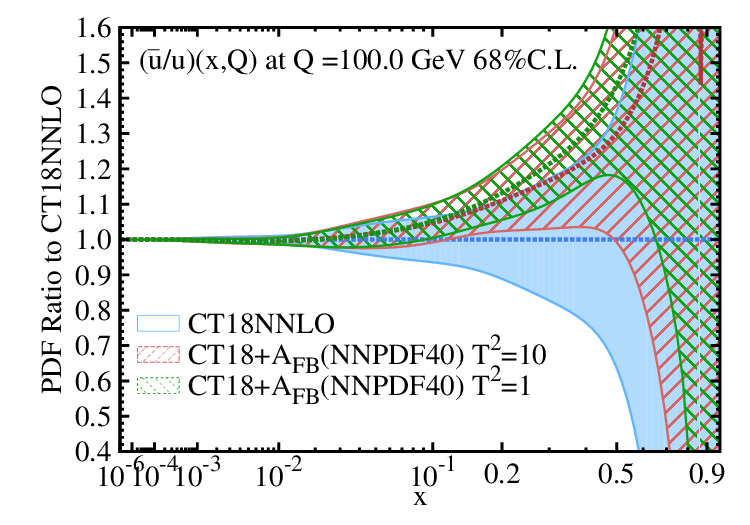}
	\includegraphics[width=0.47\textwidth]{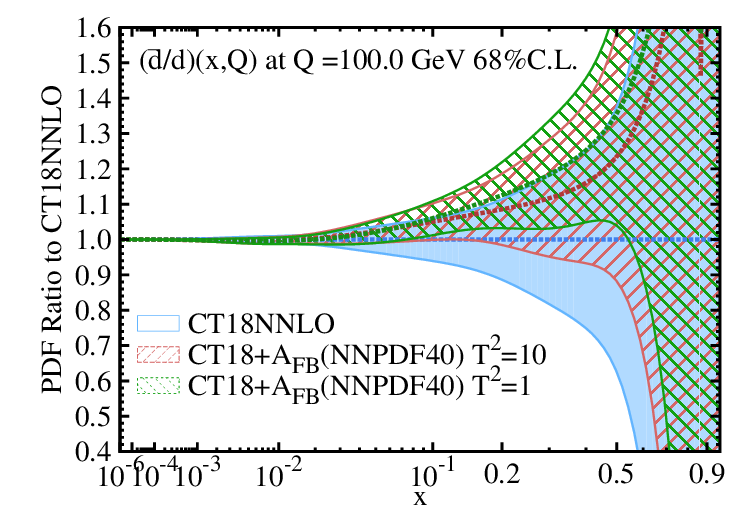}
	\caption{PDF update for $\bar{u}/u$ (left), and $\bar{d}/d$ (right) using $A_{FB}$ pseudo-data generated using NNPDF4.0. The central value and uncertainty are compared to the CT18 central value. The tolerance is set to be 10 and 1 respectively.}
	\label{fig:PDFupdate_tolerance}
\end{figure}

In this section the impact of tolerance choices in the PDF updating method is discussed. In global QCD analysis, PDFs are obtained by minimizing the $\chi^2$ function. The uncertainties are defined in the relevant neighborhood of the global minimum as:
\begin{eqnarray}
	\Delta \chi^2 \le T^2,
\end{eqnarray}
where $T$ is the tolerance parameter. In the Hessian method, the PDFs are parametrized by N parameters $\left\{z_i; i=1,N \right\}$. The $\chi^2$ function can be approximated by a quadratic expansion by the parameters $z$, so that the $\chi^2$ function can be written as:
\begin{eqnarray}
	\Delta \chi^2 = T^2\sum_{i=1}^N z^2.
\end{eqnarray}
In the \texttt{{\tt ePump}} package, theory templates need to be provided as part of the PDF updating procedure. However, \texttt{{\tt ePump}} does not know what the tolerance parameter corresponds to for a given theory template. Users must provide an input of a fixed tolerance parameter $T$, or a set of dynamical tolerances $\left \{ T_i \right \}$. The tolerance parameter is used to tell the \texttt{{\tt ePump}} package how large the deviation from the global minimum is for a given theory template, which conveys the uncertainty in the new data set. 
As shown in the section~\ref{sec:update_msht}, the original PDF uncertainty as a function of $A_{FB}$ for the MSHT20 PDF is relatively small, however, this does not lead to a larger impact when we perform PDF-updating with the same pseudodata statistical precision. As long as we input a corresponding tolerance associated with the original PDF uncertainty, the final updating result will make sense.
If the tolerance parameter is smaller than the real one, it means that the PDF uncertainty of the given theory template is treated as if it deviated less from the global minimum than in reality. In other words, a smaller tolerance means that {\tt ePump} gives a larger weight to the new data set, leading to a more significant impact.
Fig.~\ref{fig:PDFupdate_tolerance} shows the PDF updating result for CT18 using psuedo-data generated by NNPDF4.0, under different tolerance choices. The result of using $T=1$, as done in the default \texttt{xFitter} profiling~\cite{Alekhin:2014irh}, gives a much stronger impact in the update compared to $T=10$, using the same pseudo-data input. Much more detailed discussions can be found in the Appendix F of~\cite{Hou:2019efy}.

For CT18 the dynamical tolerances are used in the global QCD analysis. The effective tolerance parameter is about $100$ for $90\%$ C.L., or equivalently 37 for $68\%$ C.L.. If the tolerance parameter is set to $1$ for the $68\%$ C.L., it effectively means that a weight of $37$ is given to the new data set. This will dramatically overestimate the impact of new data set.

\end{document}